\documentclass[aip,jcp,reprint,groupedaddress]{revtex4-1}

\usepackage{amsmath,amssymb}
\usepackage{graphicx}
\usepackage{bm}

\newcommand{\subscript}[1]{\ensuremath{_{\textrm{\footnotesize{#1}}}}}
\newcommand*{\ket}[1]{\left| #1 \right\rangle}

\newcommand*{\bra}[1]{\left\langle #1 \right|}

\begin{document}
\title{Monte Carlo configuration interaction applied to multipole moments, ionisation energies and electron affinities.}
\author{Jeremy P. Coe}
\author{Daniel J. Taylor}
\author{Martin J. Paterson}
\affiliation{ 
Institute of Chemical Sciences, School of Engineering and Physical Sciences, Heriot-Watt University, Edinburgh, EH14 4AS, UK.
}%

\begin{abstract}
The method of Monte Carlo configuration interaction (MCCI)\cite{MCCIGreer95,MCCIcodeGreer} is applied to the calculation of multipole moments. We look at the ground and excited state dipole moments in carbon monoxide. We then consider the dipole of NO, the quadrupole of N\subscript{2} and of  BH.  An octupole of methane is also calculated. We consider experimental geometries and also stretched bonds.  We show that these non-variational quantities may be  found to relatively good accuracy when compared with FCI results, yet using only a small fraction of the full configuration interaction space.  MCCI results in the aug-cc-pVDZ basis are seen to generally have reasonably good agreement with experiment.  We also investigate the performance of MCCI when applied to ionisation energies and electron affinities of atoms in an aug-cc-pVQZ basis. We compare the MCCI results with full configuration-interaction quantum Monte Carlo\cite{AlaviFCIQMCIonise,AlaviFCIQMCAffinity} and `exact' non-relativistic results.\cite{AlaviFCIQMCIonise,AlaviFCIQMCAffinity}  We show that MCCI could be a useful alternative for the calculation of atomic ionisation energies however electron affinities appear much more challenging for MCCI.  Due to the small magnitude of the electron affinities their percentage errors can be high, but with regards to absolute errors MCCI performs similarly for ionisation energies and electron affinities.
\end{abstract}

\maketitle

\section*{Introduction} 

Monte Carlo configuration interaction (MCCI), created by the group of J. C. Greer,\cite{MCCIGreer95,MCCIcodeGreer} attempts to produce a compact wavefunction that can be close in accuracy to the full configuration interaction (FCI). This procedure exploits the observation that, in many systems, numerous states in a FCI contribute almost nothing to the wavefunction.  In this general approach it is not alone: a number of methods have been proposed which often essentially aim to discover the states necessary for a good description of the system without performing a FCI, see, e.g., Ref.~\citenum{NovelTruncatedCI} for a review.  MCCI offers the possibility of recovering much of the static and dynamic correlation using only a very small fraction of the configurations required for a FCI with minimal user input and, in principle, no inherent difficulties when treating excited states or multireference systems. To achieve this an iterative process of a CI calculation within a sample of coupled configurations followed by a stochastic augmentation of the sample at each step is employed.  Here configurations whose coefficient has an absolute value less than a user-specified value ($c_{\text{min}}$) in the MCCI wavefunction are eventually removed from it.  

Previous work by Greer et al has shown that single-point energies,\cite{MCCIGreer95} and bond dissociation energies\cite{dissociationGreer} for hydrogen fluoride and water, can be satisfactorily computed using this method. Electronic excitation energies for atoms have been computed using MCCI,\cite{excite1Greer} where it was found, when using an aug-cc-pVTZ basis with additional Rydberg functions, that the errors tended to be relatively small compared with experiment as were the fractions of states needed compared with a FCI.  More recently, Gy\"{o}rffy, Bartlett, and Greer\cite{GreerMCCISpectra} showed that electronic excitation energies for molecules could be calculated with errors of only around tens of meV for molecules such as nitrogen and water.  Here the MCCI wavefunctions comprised from a few thousand to around twelve thousand configuration state functions (CSFs) compared with FCI spaces of circa $10^{8}$.  The MCCI method has also been applied to ground-state potential curves in Ref.~\citenum{MCCIpotentials}.  There it was generally found to be able to produce sufficiently accurate potential energy surfaces for small molecules, even in multireference situations, using a tiny fraction of the FCI space. 

As this version of MCCI uses the magnitude of a state's coefficient in the wavefunction as the criterion for inclusion rather than a state's energy contribution we expect that properties of the exact wavefunction other than the energy should also be approximated sufficiently accurately by an MCCI wavefunction using a very small fraction of the states required for a full CI.  Here we test this idea on non-variational properties of the system: multipole moments. We calculate dipole moments for ground and excited states in carbon monoxide and the ground-state dipole moment in NO which are compared with FCI and experimental results. The quadrupole moment is calculated for N\subscript{2} and compared with experimental and FCI results.  FCI results for the quadrupole of BH and an octupole of methane are also compared with MCCI results. In addition to equilibrium geometries we also consider structures where the system may be expected to be multireference and standard methods may not work well.  The possible ability of MCCI to produce accurate enough multipoles at a range of geometries could be useful for the construction of multipole surfaces.  Finally we also consider the performance of the MCCI algorithm when applied to ionisation energies and electron affinities of atoms which we compare with FCI quantum Monte Carlo (FCIQMC)\cite{AlaviFCIQMCIonise,AlaviFCIQMCAffinity} and `exact' non-relativistic results.

\section*{Methodology}



We give a short recap of the MCCI method.\cite{MCCIGreer95,MCCIcodeGreer} MCCI stochastically adds coupled configurations to a wavefunction $\ket{\Psi_{\text{MCCI}}}=\sum_{i} c_{i} \ket{\psi_{i}}$ so that the important configurations can eventually be found regardless of their substitution level, in contrast to traditional truncation methods such as CISD, as there is no fixed reference state for MCCI. Configuration state functions (CSFs) rather than Slater determinants (SDs) are used thereby ensuring that the MCCI wavefunction is an eigenfunction of $\hat{S}^{2}$ and producing a wavefunction with fewer states.  However, the construction of linearly independent CSFs and the Hamiltonian matrix when using CSFs is more computationally demanding. An outline of the MCCI algorithm is below:
\begin{enumerate}
\item{Randomly augment the current MCCI wavefunction with single and double substitutions.}
\item{Construct the Hamiltonian matrix and diagonalize.}
\item{Remove new states whose coefficient is lower in magnitude than $c_{\text{min}}$ (pruning).}
\item{Every $10$ iterations remove all states with coefficients lower in magnitude than $c_{\text{min}}$ (full pruning).}
\item{Return to Step $1$.}  
\end{enumerate}

 We note that current states with coefficient greater than a certain value will always have single or double substitutions attempted from them while other states have a $50\%$ chance of this occurring. There is no augmentation or removal of states on the last iteration, but on the penultimate iteration all states with coefficients lower in magnitude than $c_{\text{min}}$ are removed. Furthermore the program can run in parallel with newly discovered and retained states broadcast to all other processors.  The states comprising the current MCCI wavefunction are stored at each step thereby allowing a calculation to be restarted using a previous wavefunction as the initial guess but with a smaller $c_{\text{min}}$ if necessary.  This means that if the accuracy is not sufficient at one $c_{\text{min}}$ then the calculation can be improved more efficiently than if it were just run again at a lower $c_{\text{min}}$ starting from the Hartree-Fock (HF) reference. We attempt to run the MCCI calculation for enough time so that the property of interest appears to have essentially converged over a number of iterations. We acknowledge that due to the random nature of the procedure there is always a small chance that further iterations may produce a change in the calculated property.  The diagonalization using the Davidson algorithm\cite{Davidson} is the rate limiting step when considering systems whose FCI space is large. The MCCI wavefunction is thus currently restricted to a maximum of around $10^{5}$ CSFs.  For the pruning step we use wavefunction normalisation.  For the multipole calculations this uses the coefficients after diagonalization as in the original program.\cite{MCCIcodeGreer} For the ionisation energy and electron affinity calculations we try to give a more balanced treatment of the atom and its ion by using the MCCI pruning method of Ref.~\citenum{GreerMCCISpectra} to approximate an orthogonal CSF basis.

We use a modified version of the MCCI program for the results in this work. Occupied HF molecular orbitals are used to construct the initial MCCI wavefunction and, unless otherwise stated, all electrons are correlated. For the multipole moment calculations we generate the molecular orbital integrals using the program Columbus\cite{Columbus} while we use MOLPRO\cite{MOLPRO} to calculate the molecular orbital integrals for the ionisation and electron affinity results.  For the FCI energy and multipole calculations we use PSI3.\cite{PSI3}

\section*{RESULTS}


\section*{Dipole moment results}
 

\subsection*{Carbon Monoxide}

The dipole moment in atomic units of a linear molecule oriented along the z axis may be calculated as
\begin{equation}
\mu=-\bra{\Psi}\hat{z}\ket{\Psi}+\sum_{i}z_{i}Q_{i}.
\end{equation}
Here $Q_{i}$ is the nuclear charge of atom $i$.
The ground-state dipole moment of carbon monoxide, although fairly small, when calculated using HF strikingly has the incorrect sign compared with experimental results.  Previous work has suggested that the accuracy of the dipole calculation depends on the amount of correlation accounted for.\cite{Scuseria91COdipole} The bond length ($2.1316$ Bohr)  and the experimental dipole value ($0.122$ Debye) are taken from Ref.~\citenum{COdipoleExperiment}.  The positive value for the dipole here signifies a polarity of $C^{-}O^{+}$.

\begin{figure}[h!]\centering
\includegraphics[width=.45\textwidth]{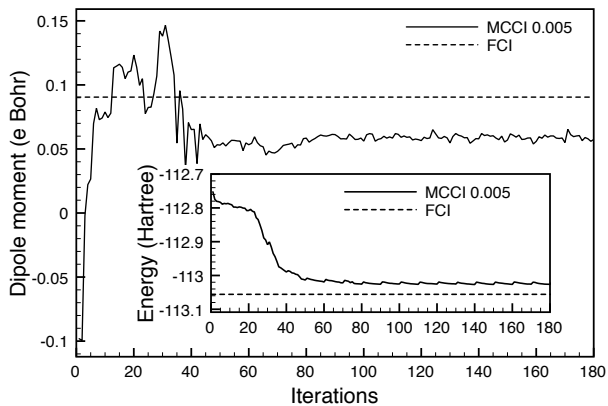}
\caption{MCCI results with $c_{\text{min}}=5\times10^{-3}$ for the dipole moment (e Bohr) against iteration number and FCI result for CO using the cc-pVDZ basis set with two frozen core orbitals. Adapted from  Ref.~\citenum{NovelTruncatedCI}. Inset: Energy (Hartree) against iteration number.}\label{fig:COdipoleVDZ1}
\end{figure}

With a cc-pVDZ basis, two frozen core orbitals and a cut-off value of $c_{\text{min}}=5\times10^{-3}$, we see in Fig.~\ref{fig:COdipoleVDZ1} that the MCCI method, starting from close to the incorrectly signed result of the HF single SD, quickly reaches a correctly signed value which converges at around half of the FCI value.  The non-variational nature can clearly be seen in Fig.~\ref{fig:COdipoleVDZ1} as it is initially far below its converged value then quickly overshoots it.  This value used only $833$ CSFs compared with a FCI space, with spatial symmetry considerations, of around $10^{9}$ SDs.

\begin{figure}[h!]\centering
\includegraphics[width=.45\textwidth]{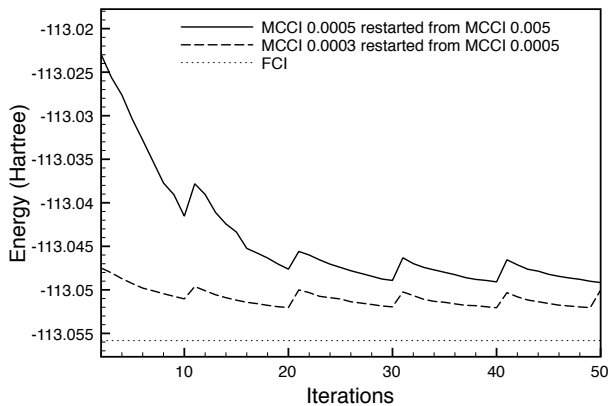}
\caption{MCCI and FCI energy (Hartree) against iteration number for CO using the cc-pVDZ basis set with two frozen core orbitals.}\label{fig:COenergyVDZ2}
\end{figure}

We calculated the FCI energy ($-113.05583$ Hartree) and dipole moment ($0.23$ D) using PSI3\cite{PSI3} for comparison and the convergence of the MCCI energy towards these values is displayed in the inset and main part of Fig.~\ref{fig:COdipoleVDZ1}.  We note that we have only recovered $88\%$ of the correlation energy when using $c_{\text{min}}=5\times10^{-3}$. When we increase the accuracy of the correlation energy by lowering $c_{\text{min}}$ we can achieve $98.1\%$ of the correlation with around $4\times10^{4}$ CSFs when using $c_{\text{min}}=3\times10^{-4}$ (see Fig.~\ref{fig:COenergyVDZ2}). The periodic behaviour of the MCCI energy at convergence is apparent in Fig.~\ref{fig:COenergyVDZ2}, and this is due to the full pruning step every ten iterations causing a small increase in the energy when a number of states are removed. New states are then added and some kept. Although their addition may have lowered the coefficients of some of the original states so that they are now below the threshold for retention, these original states are not checked for removal until the next full pruning step.  Hence as the energy is variational it lowers as more states are added until ten iterations later when all states are again considered for deletion.  This periodic behaviour is indicative of the energy calculation essentially converging.

\begin{figure}[h!]\centering
\includegraphics[width=.45\textwidth]{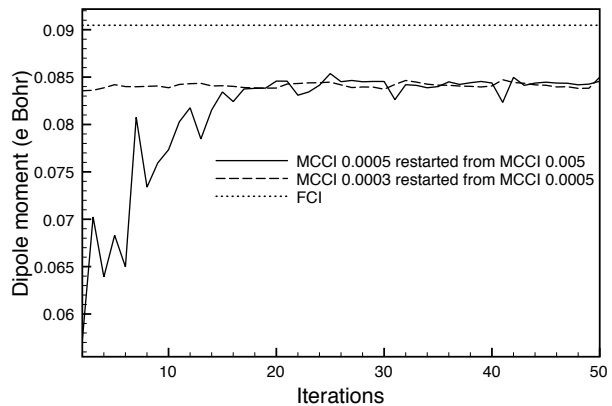}
\caption{MCCI results for the dipole moment (e Bohr) against iteration number for CO using the cc-pVDZ basis set with two frozen core orbitals.  Adapted from Ref.~\citenum{NovelTruncatedCI}.}\label{fig:COdipoleVDZ2}
\end{figure}

In Fig.~\ref{fig:COdipoleVDZ2} we see that the dipole moment is also closer to the FCI results as the cut-off value is reduced.  In these calculations the wavefunction from a previous, larger cut-off, computation has been employed as the initial wavefunction and the procedure restarted.   

\begin{figure}[h!]\centering
\includegraphics[width=.45\textwidth]{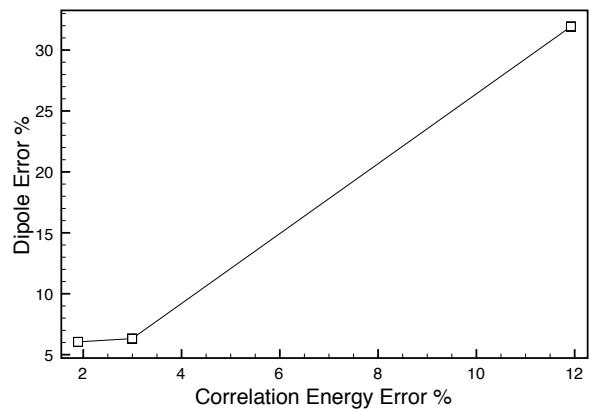}
\caption{MCCI percentage errors when compared with the FCI. Dipole percentage error plotted against correlation energy percentage error for CO using the cc-pVDZ basis set with two frozen core orbitals for three $c_{\text{min}}$ values ($5\times10^{-3}$, $5\times10^{-4}$ and $3\times10^{-4}$).  Here decreasing $c_{\text{min}}$ corresponds to decreasing correlation energy percentage error. }\label{fig:COdipoleError}
\end{figure}

In Fig.~\ref{fig:COdipoleError} we display percentage errors when comparing the MCCI results to those of the FCI.  The dipole percentage error is plotted against the correlation energy percentage error for the three cut-off values considered ($5\times10^{-3}$, $5\times10^{-4}$ and $3\times10^{-4}$) where a decreasing cut-off corresponds to a decrease in the correlation energy error.  Here we see that although the dipole error is somewhat larger it appears to decrease with decreasing correlation energy error.

\begin{figure}[h!]\centering
\includegraphics[width=.45\textwidth]{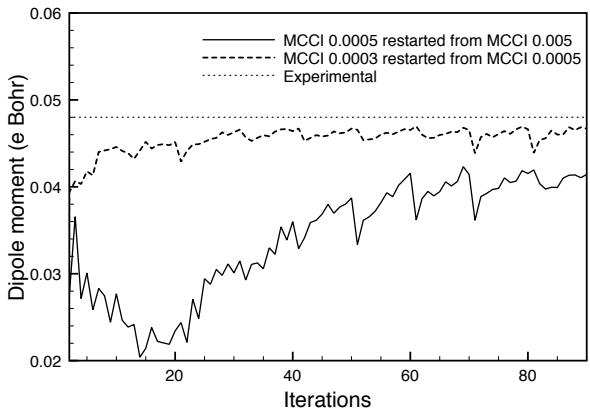}
\caption{MCCI results for the dipole moment (e Bohr) against iteration number for CO when using the aug-cc-pVDZ basis set.  Adapted from Ref.~\citenum{NovelTruncatedCI}.}\label{fig:COdipoleAUGVDZ}
\end{figure}

We note that the FCI dipole in cc-pVDZ basis has a large percentage error compared with the experimental result although the absolute error is only about $0.1D$. Diffuse functions would be expected to be important for the correct calculation of multipoles as a better description of the wavefunction further away from the atom may be needed. Hence we also considered the aug-cc-pVDZ basis with no frozen orbitals.  In this case the calculation is far beyond a FCI.  The results are depicted in Fig.~\ref{fig:COdipoleAUGVDZ} and we find that a good agreement with experiment is found as we reduce $c_{\text{min}}$ to $3\times10^{-4}$ to give a dipole moment of $0.11$ Debye.  This used $55,913$ CSFs compared with a FCI space, without spatial symmetry considerations, of around $10^{15}$ SDs.

We acknowledge that other methods such as CCSD will be more efficient to calculate the dipole of this ground state at an equilibrium geometry. For example the CCSD non-relaxed dipole can be calculated very quickly for cc-pVDZ and gives $0.0996$ e Bohr.  However we now apply MCCI to a geometry where CCSD performs poorly and then excited states.

\subsubsection*{Stretched bond length}

\begin{figure}[h!]\centering
\includegraphics[width=.45\textwidth]{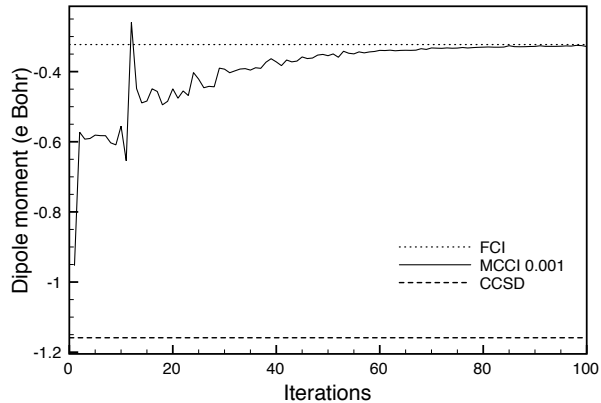}
\caption{MCCI, FCI and CCSD results for the dipole moment (e Bohr) against iteration number for CO using the cc-pVDZ basis set with two frozen core orbitals at the stretched geometry of bond length $R=4$ Bohr.}\label{fig:COR=4}
\end{figure}

Carbon monoxide at a bond length of $R=4$ Bohr was much more challenging for FCI and we note that the RMS for the error in the CI vector was $3\times10^{-2}$ in PSI3,\cite{PSI3} when taken close to the limits possible with our hardware, compared with a default requirement of $10^{-4}$. The CCSD non-relaxed dipole was calculated with MOLPRO\cite{MOLPRO} as $-1.16$ e Bohr and we note that numerical derivatives using central differences and a step size of $10^{-4}$ in field strength gave $-1.17$ for CCSD and $-1.31$ for CCSD(T).  In Fig.~\ref{fig:COR=4} we plot the CCSD non-relaxed dipole and the dipole calculated with FCI and MCCI.  We see that the MCCI calculation rapidly moves towards the FCI result and the final MCCI wavefunction gives a dipole that is difficult to distinguish from the FCI result on the scale of the graph.  The final MCCI wavefunction used $12,669$ CSFs compared with the FCI space of around $10^{9}$ SDs.  The system is strongly multireference here as the largest nine FCI coefficients have absolute values between $0.24$ and $0.30$.  Methods based on a single-reference would be expected to struggle here and we indeed observe this for CCSD and CCSD(T).  MCCI in principle has no inherent problems when dealing with multireference systems and the result here suggests that it can work well for the calculation of a multipole moment, as well as the energy, for such a system.

\subsubsection*{Triplet state}

\begin{figure}[h!]\centering
\includegraphics[width=.45\textwidth]{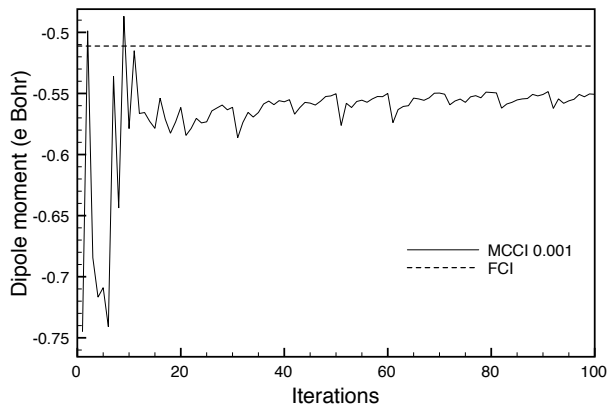}
\caption{MCCI results for the dipole moment (e Bohr) of the first triplet state against iteration number for CO at $2.278$ Bohr when using the cc-pVDZ basis set with two frozen cores and $c_{\text{min}}=10^{-3}$ compared with the FCI result.}\label{fig:COTripletVDZ}
\end{figure}

We now consider the first triplet state $^{3}\Pi$ using the experimental bond length of $2.278$ Bohr.\cite{COTripletLengthCite}  We plot the dipole moment versus iteration in Fig.~\ref{fig:COTripletVDZ} and note that now the dipole points in the opposite direction to the ground singlet state and again the non-variational nature is apparent.  The MCCI result, with $c_{\text{min}}=10^{-3}$, is in fairly good agreement with the FCI result and used $5,447$ CSFs compared with a FCI space of $8.6\times 10^{8}$ SDs.

The calculated dipole using MCCI with $c_{\text{min}}=10^{-3}$ and an aug-cc-pVDZ basis with no frozen cores gives $-1.584$ Debye with $7047$ CSFs and is in reasonable agreement with experiment ($-1.3740$ Debye).\cite{COdipoleTripletExperiment}  The agreement is better at a cut-off of $5\times 10^{-4}$ where $14,771$ CSFs gave $-1.49$ Debye. The FCI space consists of around  $10^{15}$ SDs without symmetry considerations here so again the MCCI results are using a very small fraction of the space.

\subsubsection*{Singlet excited states}
\begin{figure}[h!]\centering
\includegraphics[width=.45\textwidth]{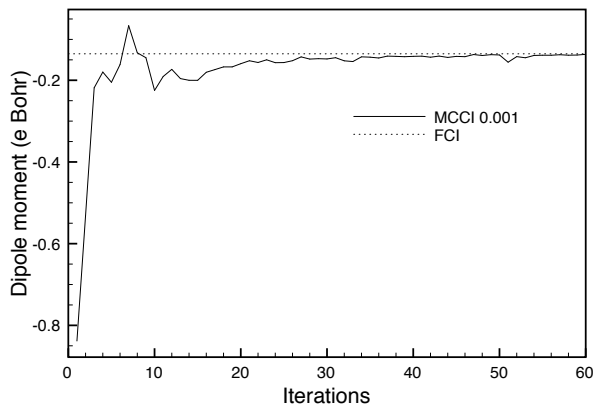}
\caption{MCCI and FCI results for the dipole moment (e Bohr) for the first $^{1}\Pi$ excited state of CO  ($B_{1}$ symmetry within $C_{2v}$)  against iteration number when using the cc-pVDZ basis set with two frozen cores at the experimental bond length of $2.334$ Bohr.}\label{fig:piCOExciteddipoleVDZ}
\end{figure}

For the first excited state $^{1}\Pi$ (ground-state of $B_{1}$ or $B_{2}$ symmetry within $C_{2v}$) in CO we consider MCCI compared with FCI results with the experimental bond length of $2.334$ Bohr as cited in Ref.~\citenum{COdipoleExciteComp}.   In Fig.~\ref{fig:piCOExciteddipoleVDZ} we see that the MCCI dipole calculation quickly converges to a value very close to the FCI on the scale of the graph. Here $10,375$ CSFs were required compared with $\sim 10^{9}$ SDs in the FCI symmetry adapted space.

When using an aug-cc-pVDZ basis with $c_{\text{min}}=10^{-3}$ and no frozen cores we find a dipole moment of  $-0.548$ Debye at the ground-state geometry using $16,487$ CSFs compared with $-0.335\pm0.013$ Debye from Ref.~\citenum{COdipoleExcitedExperiment2} while an earlier study\cite{COdipoleExcitedExperiment} found this to be $-0.15\pm0.05$ Debye. Here the signs of the experimental results have been determined by a theoretical study.\cite{COdipoleExciteComp} The result is closer to the later experiment when $c_{\text{min}}$ is lowered to $5\times10^{-4}$ to give $-0.418$ with $45,274$ CSFs.

\begin{figure}[h!]\centering
\includegraphics[width=.45\textwidth]{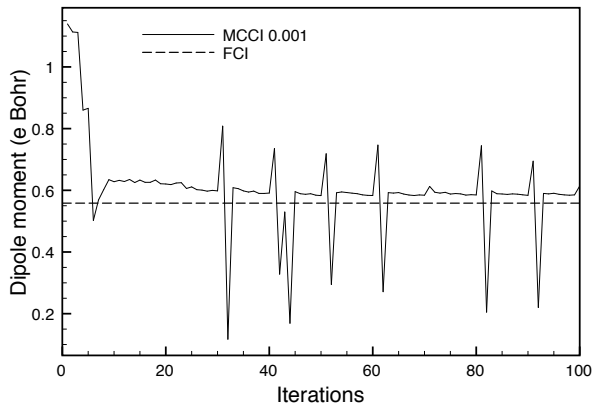}
\caption{MCCI and FCI results for the dipole moment (e Bohr) for the first  ($^{1}\Sigma^{+}$) excited state of CO at $2.116$ Bohr ($A_{1}$ symmetry within $C_{2v}$)  against iteration number when using the cc-pVDZ basis set with two frozen cores .}\label{fig:COExciteddipoleVDZ}
\end{figure}

We now consider the first excited state of $A_{1}$ symmetry within $C_{2v}$ ($^{1}\Sigma^{+}$) in CO and use the experimental bond length cited in Ref.~\citenum{COdipoleExciteComp} of $2.116$ Bohr.  We see in Fig.~\ref{fig:COExciteddipoleVDZ} that the first excited state of $A_{1}$ symmetry dipole calculation with the cc-pVDZ basis quickly approaches its converged value after starting with a too high but same signed value. The stable value is close to the FCI result.  We note that the calculation of this excited state is not as stable in that oscillations occasionally occur.  The energy sometimes rises sharply after a full pruning step here and this is accompanied by an increase then decrease in the dipole before it returns to essentially its almost converged value.  It seems that sometimes states that are important for this system are removed during a full prune.  The energy appears to recover almost to its previous value in one iteration, but it appears to take at least two iterations for the dipole moment and its non-variational nature is apparent. This more sensitive behaviour to the removal of states may be connected to the level of cut-off and the use of the second eigenvalue from the MCCI diagonalization routine.  This used $8,988$ CSFs for the final MCCI wavefunction compared with the symmetry adapted FCI space of circa $10^{9}$ SDs.

An experimental study\cite{COdipoleExcitedExperiment} found this dipole to be $1.60\pm 0.15$ Debye while a later work\cite{COdipoleExcitedExperiment2} found it to be $1.95\pm0.03$ Debye.  The sign was not determined in these experiments but Ref.~\citenum{COdipoleExciteComp} found the dipole of this excited state to be around $-2.79$ Debye at the ground-state geometry using MCSCF and CIS with about $27,000$ CSFs, while that of the ground-state was calculated as $0.32$ Debye.  When using an aug-cc-pVDZ basis we seem to find an essentially converged value of $1.762$ Debye  with $22,198$ CSFS when using $c_{\text{min}}= 10^{-3}$. A value of  $1.69$ Debye with $71,857$ CSFs was found using $c_{\text{min}}= 5\times 10^{-4}$ but here the last value was an oscillation so we used the second last iteration where all states were considered for removal.  However there were fewer oscillations when using this basis.  The values are reasonably near to the earlier experimental work and become more similar as $c_{\text{min}}$ is decreased but the non-variational nature  and aug-cc-pVDZ basis could be responsible for this. However, the sign is different to that of the computational study of  Ref.~\citenum{COdipoleExciteComp} as we find that the dipole is in the same direction as that of the ground state, but  we note that EOM-CCSD calculations using MOLPRO\cite{MOLPRO} are in agreement with the sign and magnitude of the MCCI results as it gives a dipole of about $1.60$ Debye for cc-pVDZ and  $1.72$ Debye for aug-cc-pVDZ both with two frozen cores.

\subsection*{NO}
The dipole of NO in its doublet ground-state has been measured as $0.157$ Debye\cite{NOdipoleExperiment} and its sign verified as positive in Ref.~\citenum{NOdipolesign} corresponding to $N^{-}O^{+}$. We use the experimental bond length of $1.1508$ angstroms cited in Ref.~\citenum{NOdipoleComp} and, in addition to MCCI together with FCI results, we calculate the UCCSD dipole moment using the numerical derivative of the energy with respect to the electric field in MOLPRO.\cite{MOLPRO}  To achieve this we use central differences and a step size of $10^{-4}$ in the field strength.

\begin{figure}[h!]\centering
\includegraphics[width=.45\textwidth]{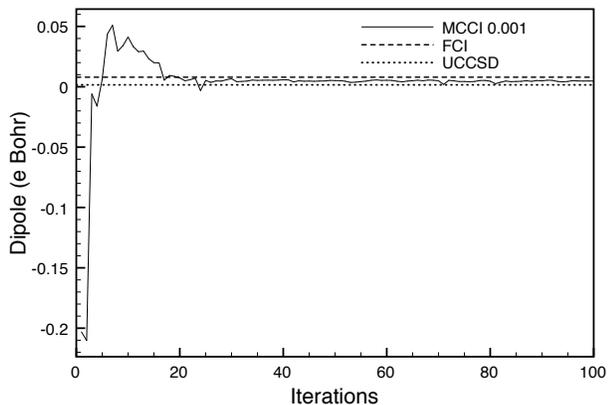}
\caption{MCCI, UCCSD and FCI results for the dipole moment (e Bohr) against iteration number for NO when using the 6-31G basis set.}\label{fig:NOdipole631G}
\end{figure}

Fig.~\ref{fig:NOdipole631G} shows that with a 6-31G basis MCCI quickly recovers the correct sign after starting with a value close to the incorrect Hartree-Fock dipole, and gives a reasonable dipole moment ($0.0048$ e Bohr), at this level of cut-off, in comparison with the FCI result ($0.0079$ e Bohr)  where it is more accurate than UCCSD ($0.0016$ e Bohr) but the absolute differences in accuracy are very small.

Here there are around $3\times 10^{8}$ SDs in the FCI space when considering symmetry compared with $3,274$ CSFs for MCCI with no cores frozen in both cases.

\begin{figure}[h!]\centering
\includegraphics[width=.45\textwidth]{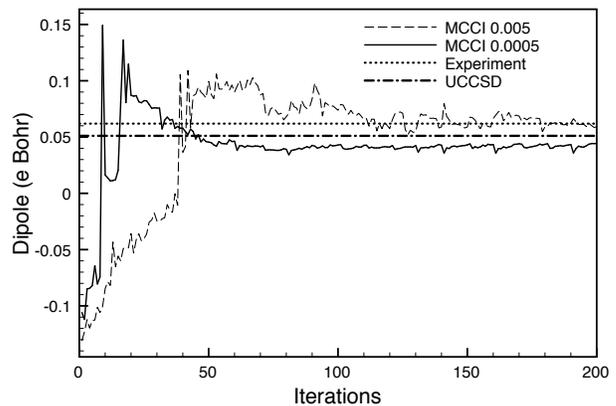}
\caption{MCCI and UCCSD results for the dipole moment (e Bohr) against iteration number for NO when using the aug-cc-pVDZ basis set compared with experiment.}\label{fig:NOdipoleAUGVDZ}
\end{figure}

In Fig.~\ref{fig:NOdipoleAUGVDZ} we see that with an aug-cc-pVDZ basis the MCCI result with the larger cut-off is close to that of experiment while UCCSD is just a little lower. However when the accuracy of MCCI is increased by lowering $c_{\text{min}}$ to $5\times10^{-4}$ the calculated dipole is below that of UCCSD suggesting that perhaps the most accurate result in this basis would be below experiment, but the non-variational nature means this prediction is not certain.  The dipole of around $0.12$ Debye at the highest accuracy MCCI considered is still in fairly good agreement with experiment and we note that this used $17,188$ CSFs compared with a Full CI space, without symmetry considerations, of circa $10^{16}$ Slater determinants.

\section*{Quadrupole moments}

\subsection*{Nitrogen molecule}
The Buckingham traceless quadrupole moment tensor\cite{BuckinghamQuad} is defined as
\begin{equation}
Q_{\alpha\beta}=\frac{1}{2}\sum_{i} q_{i} (3r_{i\alpha}r_{i\beta}-\delta_{\alpha\beta}r_{i}^{2})
\end{equation}
where $r=(x,y,z)$.
For a diatomic molecule, aligned along the z axis, with its centre of mass at the origin this becomes for $Q_{zz}$
\begin{equation}
Q_{zz}=\frac{1}{2}(\bra{\Psi} \hat{x}^{2} + \hat{y}^{2}-2\hat{z}^{2}  \ket{\Psi} +2Z_{A}R^{2}_{A0}+2Z_{B}R^{2}_{0B}).
\end{equation}
Here $Z_{i}$ is the charge of nucleus $i$ and $R_{i0}$ is the distance between nucleus $i$ and the origin.

For N\subscript{2} the traceless quadrupole moment with respect to the centre of mass at the origin has been measured\cite{N2quadexp} as $(-4.65 \pm 0.08) \times 10^{-40}$ Cm$^{2}$ and revised in a theoretical work\cite{NitrogenQuadTheory} using an improved value for the correction term to give  $(-5.01 \pm 0.08) \times 10^{-40}$ Cm$^{2}$. We use the latter value and the experimental bond length of $2.07432$ Bohr cited in Ref.~\citenum{NitrogenQuadTheory}. 

\begin{figure}[h!]\centering
\includegraphics[width=.45\textwidth]{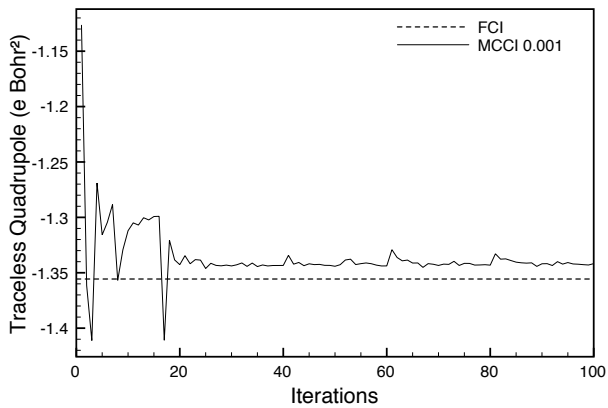}
\caption{MCCI and FCI results for the traceless quadrupole moment $Q_{zz}$ (e Bohr$^{2}$) against iteration number for N\subscript{2} when using the cc-pVDZ basis set with two frozen cores.}\label{fig:quadN2VDZ}
\end{figure}

With a cc-pVDZ basis and two frozen cores the cutoff of $10^{-3}$ gives reasonable agreement with the FCI results (see Fig.~\ref{fig:quadN2VDZ}). FCI results were calculated with a modified version of PSI3.\cite{PSI3}  The MCCI result used 5761 CSFs compared with the SD space, when considering symmetries, of $5.4\times 10^{8}$

\begin{figure}[h!]\centering
\includegraphics[width=.45\textwidth]{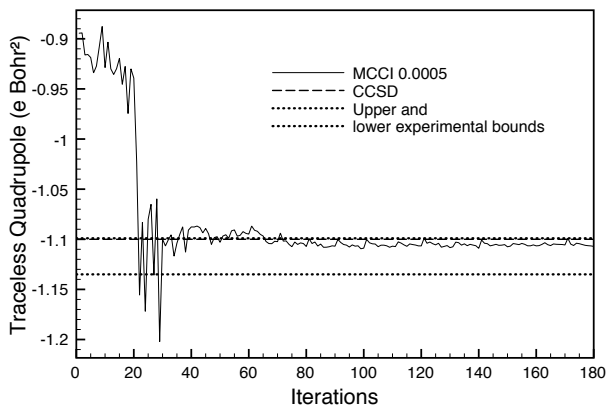}
\caption{MCCI and CCSD results for the traceless quadrupole moment $Q_{zz}$ (e Bohr$^{2}$) against iteration number for N\subscript{2} when using the aug-cc-pVDZ basis set.   Adapted from Ref.~\citenum{NovelTruncatedCI}.}\label{fig:quadN2augVDZ}
\end{figure}

  With the aug-cc-pVDZ basis the results using $5\times 10^{-3}$ were within the experimental bounds with the MCCI result a little lower than that of CCSD (see Fig.~\ref{fig:quadN2augVDZ}). The FCI space would consist of around $10^{15}$ Slater determinants if spatial symmetries are neglected while the MCCI wavefunction comprised about $22,000$ CSFs.

The MCCI result of $-1.105$ e Bohr$^{2}$ also compares favourably with CCSD(T) results calculated in Ref.~\citenum{NitrogenQuadTheory} where an aug-cc-pVDZ basis with only valence electrons correlated gave $-1.1116$  e Bohr$^{2}$ and an aug-cc-pCVQZ with all electrons correlated resulted in $-1.1159$  e Bohr$^{2}$.

\subsection*{BH}

A smaller calculation for which published FCI multipole results are available is the quadrupole of BH in an aug-cc-pCVDZ basis. We compare the MCCI results with those of FCI and coupled cluster in Ref.~\citenum{BHfciQuad}.  Here we use the experimental bond length cited in the latter paper ($2.3289$ angstroms) and the mass of the most common isotope of boron ($^{11}$B) is used to calculate the centre of mass.

\begin{figure}[h!]\centering
\includegraphics[width=.45\textwidth]{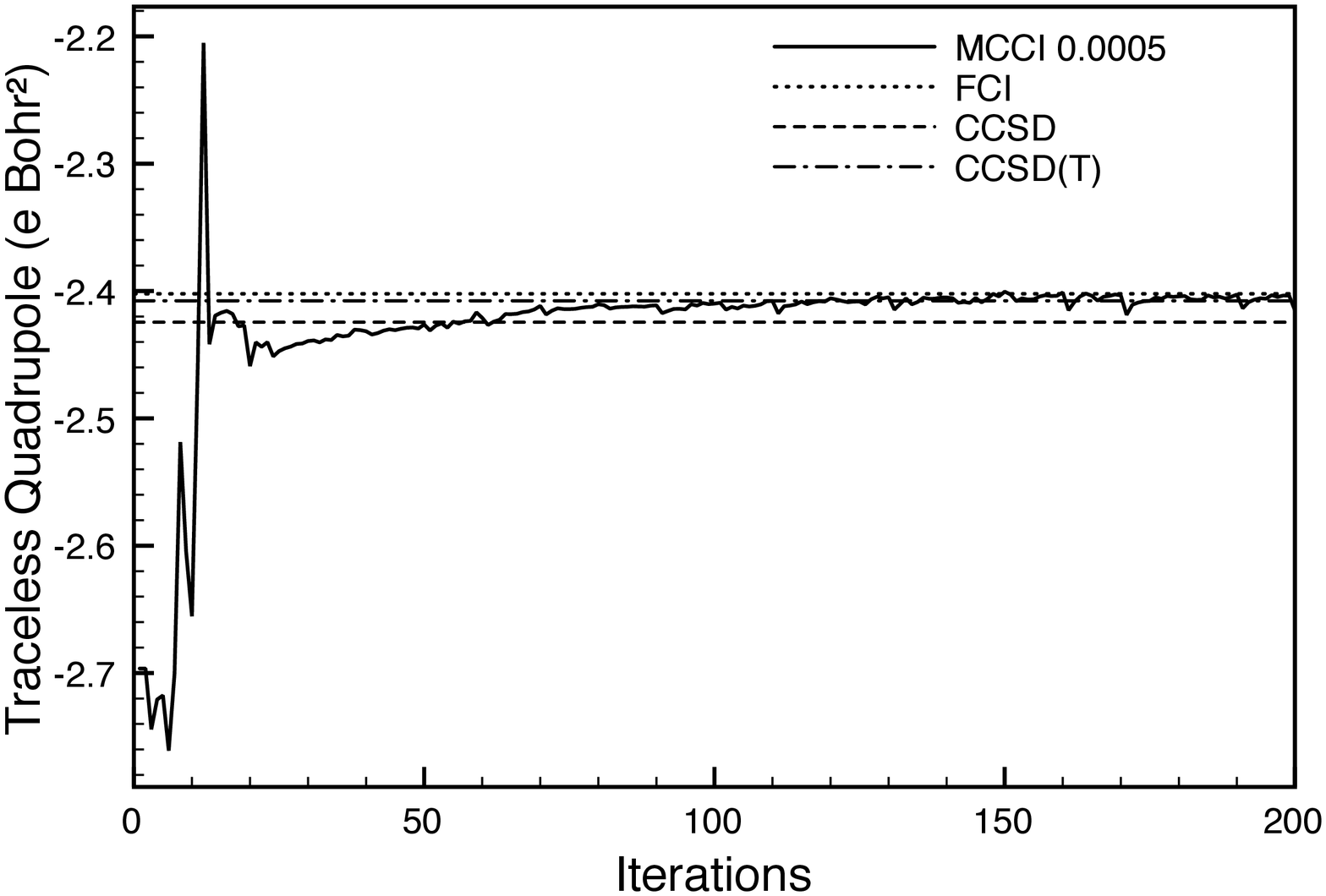}
\caption{FCI and coupled cluster results from Ref.~\citenum{BHfciQuad} and MCCI results for the traceless quadrupole moment $Q_{zz}$ (e Bohr$^{2}$) against iteration number for BH when using the aug-cc-pCVDZ basis set.}\label{fig:quadBHaugCVDZ}
\end{figure}

We see in Fig.~\ref{fig:quadBHaugCVDZ} that the quadrupole calculated using MCCI rapidly reaches a value closer to the FCI result than CCSD and is of comparable accuracy to that of CCSD(T) and FCI on the scale of the plot.  We note that the final MCCI wavefunction used $4,276$ CSFs compared with a FCI space of around $5\times 10^{7}$ SDs without symmetry considerations.

\section*{Octupole of Methane}
We calculate the octupole moment of methane at a tetrahedral geometry with an equilibrium $CH$ bond length\cite{MethaneBondExp} of $2.052$ Bohr using a cc-pVDZ basis with one frozen core.
Here we place the carbon atom at the origin and use the co-ordinates of $(x_{H},0,z_{H})$, $(-x_{H},0,z_{H})$, $(0,x_{H},-z_{H})$ and $(0,x_{H},-z_{H})$ for the hydrogen atoms where $z_{H}=1.18486$ Bohr and $x_{H}=1.67565$ Bohr.  We use the traceless octupole moment of Buckingham\cite{BuckinghamQuad} where, for our co-ordinates we have, for example,
\begin{equation}
\Omega_{xxz}=\frac{1}{2}\left( \bra{\Psi} -4\hat{x}^{2}\hat{z}+\hat{y}^{2}\hat{z}+\hat{z}^{3} \ket{\Psi}+10x_{H}^{2}z_{H}\right).
\end{equation}

\begin{figure}[h!]\centering
\includegraphics[width=.45\textwidth]{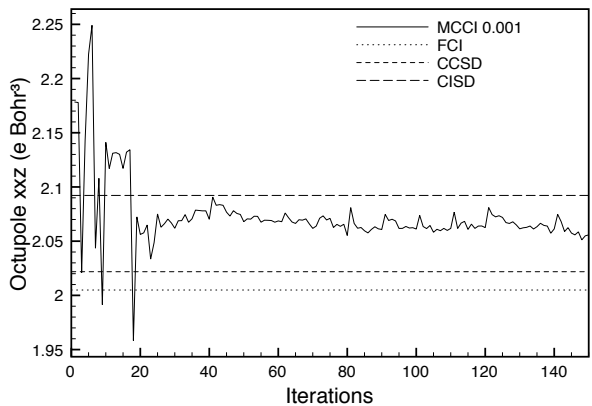}
\caption{MCCI, CISD, CCSD and FCI results for  $\Omega_{xxz}$  (e Bohr$^{3}$) of methane against iteration number when using the cc-pVDZ basis set with one frozen core.}\label{fig:MethaneOctVDZ}
\end{figure}

 We compare the MCCI value for the $\Omega_{xxz}$ component with FCI and CISD results from a modified version of PSI3\cite{PSI3} and with coupled cluster results from the program Dalton.\cite{Dalton}  Fig.~\ref{fig:MethaneOctVDZ} shows how the octupole converges relatively quickly with MCCI using $c_{\text{min}}=10^{-3}$. The value is an improvement on that of CISD but, unsurprisingly, at this equilibrium geometry CCSD is closer to the FCI result.  The MCCI wavefunction consisted of $3,330$ CSFs while the FCI space comprised of circa $4\times10^{8}$ SDs.

\subsubsection*{Stretched bond length}

\begin{figure}[h!]\centering
\includegraphics[width=.45\textwidth]{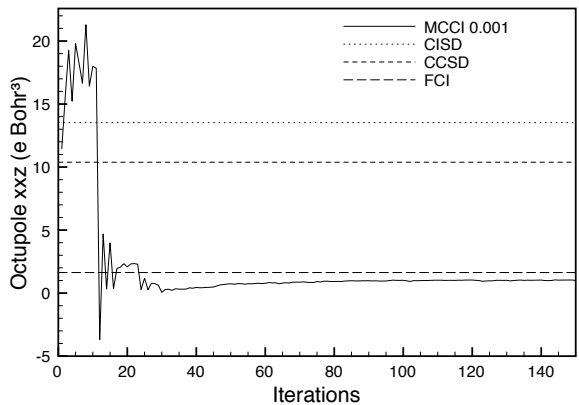}
\caption{MCCI, CISD, CCSD and FCI results for  $\Omega_{xxz}$  (e Bohr$^{3}$) of methane with $R=5$ Bohr against iteration number when using the cc-pVDZ basis set with one frozen core.}\label{fig:MethaneOctVDZlargeR}
\end{figure}

We now consider a geometry away from equilibrium of $R=5$ Bohr for all CH bonds.  This results in $x_{H}= 4.08248$ and $z_{H}= 2.88675$. Here the system is more likely to be multireference and we see from the results in Fig.~\ref{fig:MethaneOctVDZlargeR} that CISD and CCSD perform poorly giving an octupole over six times that of the FCI.  MCCI, even with a cut-off as large as $10^{-3}$, does much better, but, although the absolute difference is about $0.6$ e Bohr$^{3}$, the MCCI value is about $1.6$ times smaller compared with FCI.

\section*{Multipole Summary}
We summarise the comparison of the MCCI with the FCI multipole results calculated in this work in table \ref{tbl:multipoles}. Here the smallest cut-off MCCI results are presented and we see that over a range of states and geometries the MCCI multipoles are generally very close to the FCI results and the MCCI CSFs used are just a very small percentage of the FCI SD space. 
\begin{table*}[ht!]
\centering
\begin{tabular}{|c|c|c|c|c|c|} \hline
\textbf{Property} &\textbf{MCCI} & \textbf{FCI} & $\%$ \textbf{FCI space} &\textbf{$E_{\text{corr}}$ Error} & \textbf{Property Error}   \\ \hline
 CO $\mu$  & 0.0850  & 0.0905 &   $3.63\times10^{-3}$ & 1.89$\%$  & 6.05$\%$  \\ \hline
 CO $R=4$ $\mu$ & -0.328 & -0.323 & $1.17\times10^{-3}$& 3.79$\%$  & 1.70$\%$   \\ \hline
 CO $^{3}\Pi$ $\mu$  & -0.551   & -0.511   & $6.33\times10^{-4}$& 4.31$\%$  & 7.73$\%$    \\ \hline
 CO $^{1}\Pi$ $\mu$  &  -0.138  & -0.135   & $9.59\times10^{-4}$&  - &  2.22$\%$  \\  \hline
 CO Excited $^{1}\Sigma^{+}$ $\mu$  & 0.614    & 0.558    & $8.31\times10^{-4}$& - & 9.97$\%$  \\ \hline
 NO $\mu$ &  0.00475  & 0.00794   & $9.55\times10^{-4}$& 1.35$\%$  & 40.2$\%$  \\ \hline
 N\subscript{2} $Q_{zz}$ & -1.342   & -1.356   &  $1.07\times10^{-3}$& 1.06$\%$   & 1.02$\%$ \\ \hline
 CH\subscript{4} $\Omega_{xxz}$ &  2.056  &  2.0049  & $7.95\times10^{-4}$& 4.98$\%$  & 2.54$\%$ \\ \hline
 CH\subscript{4} $R=5$ $\Omega_{xxz}$ & 1.000   &  1.631  & $3.24\times10^{-3}$& 1.25$\%$   & 38.7$\%$  \\ \hline
\end{tabular}
\caption{Table showing MCCI and FCI multipole results in atomic units, the fraction of CSFs used in MCCI when compared with the symmetry adapted FCI space using SDs, the percentage error of the correlation energy and the percentage error of the property compared with the FCI. The cc-pVDZ basis is used except for NO which has 6-31G. No orbitals are frozen for NO, while one is frozen for CH\subscript{4} and the other results use two frozen orbitals. Experimental geometries as presented earlier in the paper are used unless otherwise stated.} \label{tbl:multipoles}
\end{table*}

The largest percentage errors are for the dipole of NO and the octupole of methane at a stretched geometry.  In the former case this is due to the dipole being very small while in the latter case the strong multireference character, suggested by the very poor performance of CCSD, may be responsible.  With the exception of these two systems the errors of the property tend to be similar to those of the correlation energy. This and the behaviour of the errors of dipole and correlation energy for CO with decreasing $c_{\text{min}}$ (Fig.~\ref{fig:COdipoleError}) hints that it may be possible to use the scheme of Ref.~\citenum{Gyorrfy2006} to approximate the value of a property, rather than the correlation energy, for $c_{\text{min}}\rightarrow 0$ through repeated MCCI calculations for various fixed numbers of CSFs.  However the non-variational nature of the properties and the large difference in correlation energy error and property error seen for stretched CH\subscript{4} suggest that caution should be used if this approach is employed in future work.

\section*{Ionisation energies}

We now use MCCI to calculate ionisation energies for atoms and compare the MCCI results with full configuration-interaction quantum Monte Carlo (FCIQMC) results\cite{AlaviFCIQMCIonise} and the `exact' non-relativistic results listed in Ref.~\citenum{AlaviFCIQMCIonise} which they extracted from Ref.~\citenum{KogaIonisation}. FCIQMC uses projector or diffusion Monte Carlo in a Slater determinant basis to approach the FCI solution without needing to diagonalise the Hamiltonian matrix. The computational difficulty for FCIQMC can be linked to the number of walkers required for convergence in the diffusion Monte Carlo calculation.

Using an aug-cc-pVQZ basis we see in Table.~\ref{tbl:ionisation} that values all within $7$ milliHartree of the FCIQMC are achieved with always fewer than an average of $4,000$ CSFs. Here we used $c_{\text{min}}=5\times10^{-4}$ and $500$ iterations on $12$ processors with the exception of the lithium cation which was still the Hartree-Fock reference after $500$ iterations so was run for $3,000$ iterations. In this case the calculation takes less than two minutes and gave a final state of $54$ CSFs.

\begin{table}[h]
\centering
\begin{tabular}{|c|c|c|c|} \hline
\textbf{Atom} & \textbf{MCCI mean CSF}s &\textbf{MCCI}  & \textbf{FCIQMC}\cite{AlaviFCIQMCIonise}   \\ \hline
 Li &  101 &197.46&  197.35 \\ \hline
 Be & 270 &341.02 & 341.89 \\ \hline
 B  &  869 &302.66 & 304.02 \\ \hline
 C  &  1,572 &411.89 & 413.10\\ \hline
 N  &  2,174 &531.42 & 535.85\\ \hline
 O  &  2,851 &491.38 & 497.35\\ \hline
 F  &  3,536 &631.46 & 638.61\\ \hline
Ne  &  3,376 &786.14 & 792.48\\ \hline
Na  &  707 &184.93 & 184.32\\ \hline
Mg  &  1,064 &275.75 &   -\\ \hline
\end{tabular}
\caption{MCCI with $c_{\text{min}}=5\times10^{-4}$ average CSFs for atom and cation. Ionisation energies in milli Hartree using aug-cc-pVQZ from MCCI and FCIQMC.\cite{AlaviFCIQMCIonise}  } \label{tbl:ionisation}
\end{table}

The sodium atom used only about $2,000$ walkers and a few minutes in Ref.~\citenum{AlaviFCIQMCIonise} compared with a FCI space of around $10^{15}$ and we find here that it requires $907$ CSFs and 142 seconds with MCCI when using $12$ processors. We can see in Table.~\ref{tbl:ionisation} that the MCCI calculation for sodium gives almost the same ionisation energy as FCIQMC. We found that the fluorine atom required the largest number of CSFs at $4,189$ compared with a FCI space of around $10^{13}$ Slater determinants and a calculation time of just over an hour using 12 processors.

\begin{figure}[h!]\centering
\includegraphics[width=.45\textwidth]{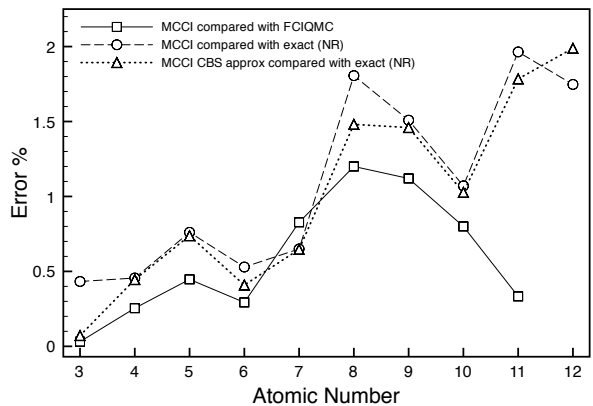}
\caption{MCCI error with  $c_{\text{min}}=5\times10^{-4}$  when compared with FCIQMC\cite{AlaviFCIQMCIonise} both using an aug-cc-pVQZ basis and `exact' non-relativistic (NR) ionisation energies.\cite{KogaIonisation} MCCI CBS approximation error with  $c_{\text{min}}=5\times10^{-4}$ compared with `exact' non-relativistic (NR) ionisation energies. }\label{fig:ionisation}
\end{figure}

We note that magnesium was not calculated in an aug-cc-pVQZ basis in Ref.~\citenum{AlaviFCIQMCIonise} due to CPU time constraints using FCIQMC  while here we note it required about $1,000$ CSFs compared with having the largest full CI space of ~$10^{17}$ and the calculation required around five minutes for the cation and less for the atom. The MCCI result gave an error of about $1.7\%$ compared with the `exact' result of $280.65$ milliHartree.  The oxygen atom was found to be particularly challenging for FCIQMC where it required $100$ million walkers and around $48$ hours on $32$ processors.  The MCCI value used $3,541$ CSFs for the atom ( $2,162$ for the cation) and required almost an hour on $12$ processors, but here the MCCI result at this level of cut-off is $8$ mHartree below that of FCIQMC although the percentage error is just $1.2$.  We see in Fig.~\ref{fig:ionisation} that this was the largest percentage error when compared with FCIQMC. This error could be brought lower but to the detriment of calculation size and time by reducing the MCCI cut-off.  It is interesting that the error rises then peaks at oxygen when compared with FCIQMC but the error with the `exact' gives the impression of an overall trend for a rising error. Although oxygen now has the second largest error with sodium the largest.  We note that all the MCCI errors are under $2\%$ when compared with the `exact' results.

We also consider approximating the MCCI ionisation energy in the complete basis set (CBS) limit.  While the MCCI ionisation energy is not variational and would not be expected to behave monotonically with increasing basis size, the underlying energies should smoothly approach the CBS limit. We use the scheme of Ref.~\citenum{KartonMartin}, given by $E_{x}=E_{\infty}+A(x+1)e^{-9\sqrt{x}}$, to approximate the CBS limit for the Hartree Fock energy. For the MCCI correlation energy we use $E_{corr,x}=E_{corr,\infty}+Bx^{-3}$ from Ref.~\citenum{HelgakerScheme} to approximate the CBS limit.  Here $x=2$ for aug-cc-pVDZ, $x=3$ for aug-cc-pVTZ and so on. We fit the schemes to the results at aug-cc-pVTZ and aug-cc-pVQZ.   We note that we neglect aug-cc-pVDZ as it is often considered better to fit to two points rather than three if the third is thought to be too far from the CBS limit. Furthermore the use of aug-cc-pVTZ and aug-cc-pVQZ for the HF extrapolation was found to not work so well in Ref.~\citenum{KartonMartin}, but we note that, when using the schemes here, the change in the HF energy is much smaller than the change in the correlation energies.  In Fig.~\ref{fig:ionisation} we display the error of the MCCI approximate CBS when compared with the `exact' ionisation energies.  We see that the approximation to the MCCI CBS at $c_{\text{min}}=5\times10^{-4}$ has a lower percentage error except in the case of the result for magnesium. In general, the approximate CBS percentage errors tend not to be substantially lower but the results for lithium and oxygen are noticeably more accurate. 

\begin{figure}[h!]\centering
\includegraphics[width=.45\textwidth]{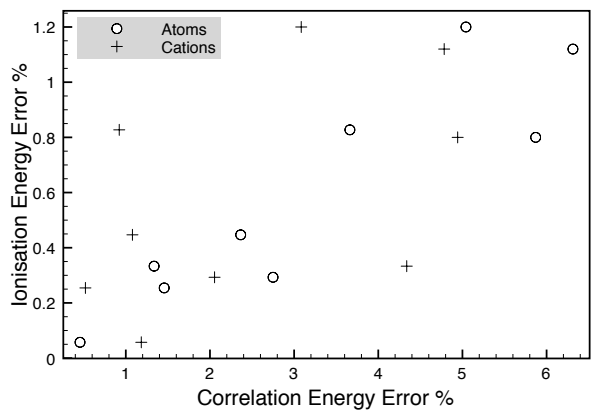}
\caption{Percentage error in the ionisation energy plotted against the percentage error in the correlation energy of the atom (circles) or cation (crosses). All results are for MCCI at $c_{\text{min}}=5\times10^{-4}$ compared with FCIQMC\cite{AlaviFCIQMCIonise} both using an aug-cc-pVQZ basis. }\label{fig:ionisationErrorvEcorr}
\end{figure}

We see in Fig.~\ref{fig:ionisationErrorvEcorr} that for a given percentage error in the ionisation energy the percentage errors in the correlation energy are fairly similar for the atom and cation when comparing MCCI and FCIQMC results.  We note that the cation usually, but not always, has a slightly lower error for the correlation energy.  The general trend is for the ionisation energy error to increase with increasing correlation energy error with the former generally smaller than the latter.  There appears to be a much stronger linear relationship in the percentage errors for the atoms than for the cations: the statistical correlation between the results is $0.91$ for the atoms and $0.51$ for the cations.

\section*{Electron Affinities}

We finally compare electron affinities calculated with MCCI with those of initiator FCIQMC (i-FCIQMC)\cite{AlaviFCIQMCAffinity} and `exact' non-relativistic results from Ref.~\citenum{AlaviFCIQMCAffinity} which are again extracted from Ref.~\citenum{KogaIonisation}. Electron affinities are considered computationally difficult, partly due to the requirement of achieving a balanced error in the atom and anion calculations with the latter's extra electron often weakly bound. We ran the calculation for $3000$ iterations for $c_{\text{min}}=5\times10^{-4}$ but with the exception of sodium the percentage errors were not much different to the values at $500$ iterations.  The largest FCI space was sodium at around $10^{17}$ for which the MCCI calculation of the anion required $864$ CSFs and less than ten minutes.  The largest number of CSFs at $c_{\text{min}}=5\times10^{-4}$ was $6,054$ for the oxygen anion compared with a FCI space of around $10^{13}$. This calculation at $3,000$ iterations needed around $19$ hours on twelve processors but we note that the results were not much different to $500$ iterations which required less than $3$ hours.

\begin{figure}[h!]\centering
\includegraphics[width=.45\textwidth]{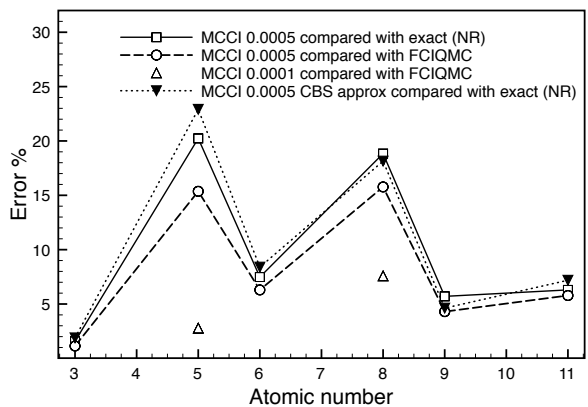}
\caption{MCCI error with  $c_{\text{min}}=5\times10^{-4}$  when compared with i-FCIQMC\cite{AlaviFCIQMCAffinity} both using an aug-cc-pVQZ basis and `exact' non-relativistic (NR) electron affinities.\cite{KogaIonisation} MCCI CBS aprroximation error with $c_{\text{min}}=5\times10^{-4}$ when compared with `exact' (NR) electron affinities.}\label{fig:affinities}
\end{figure}

We see in Table.~\ref{tbl:affinities} that the MCCI values are reasonably close to the i-FCIQMC results with the difference always less than $10$ milliHartree and the number of mean CSFs always fewer than $5,000$. However due to the electron affinities being much smaller than the ionisation values the percentage errors seen in Fig.~\ref{fig:affinities} are much higher than the ionisation errors when using $c_{\text{min}}=5\times10^{-4}$.  Particularly we see in Table.~\ref{tbl:affinities} that the absolute error in boron is actually quite low, but the very small electron affinity means the percentage error is large, while oxygen has the largest absolute and largest percentage error when compared with the i-FCIQMC results.

\begin{table}[h]
\centering
\begin{tabular}{|c|c|c|c|}\hline
\textbf{Atom} & \textbf{MCCI mean CSFs} & \textbf{MCCI}  & \textbf{FCIQMC}\cite{AlaviFCIQMCIonise}   \\ \hline
 Li &  $295$ &22.34&  22.60 \\ \hline
 B  &  $2,362$ &8.18 & 9.67 \\ \hline
 C  &  $3,055$ &43.20 & 46.10\\ \hline
 O  &  $4,869$ &43.93 & 52.15\\ \hline
 F  &  $4,538$ &118.95 & 124.29 \\ \hline
Na  &  $909$ &18.87 & 20.03\\ \hline
\end{tabular}
\caption{MCCI with $c_{\text{min}}=5\times10^{-4}$ average CSFs for atom and anion. Electron affinities in milliHartree using aug-cc-pVQZ from MCCI and i-FCIQMC.\cite{AlaviFCIQMCAffinity}} \label{tbl:affinities}
\end{table}

 In Fig.~\ref{fig:affinities} with $c_{\text{min}}=5\times10^{-4}$ the largest MCCI error is now around $20\%$ when compared with the `exact' results and $15\%$ when compared with i-FCIQMC.  The two most difficult systems are boron and oxygen, which, by reducing $c_{\text{min}}$ to $10^{-4}$ we can get their errors with MCCI to around $3\%$ and $8\%$ respectively when compared with i-FCIQMC. However this comes at a computational cost:  for this cut-off the boron anion needed $13,734$ CSFs while the oxygen anion calculation required $37,225$ CSFs and a calculation time of $28$ hours on $8$ processors when the reference state was the MCCI wavefunction from the $c_{\text{min}}=5\times10^{-4}$ calculation.

We also approximate the CBS limit of the MCCI electron affinities using the same procedure as for the MCCI ionisation energies. With $c_{\text{min}}=5\times10^{-4}$ we see in Fig.~\ref{fig:affinities} that the MCCI electron affinities when approximating the CBS have an error that is similar to the results using an aug-cc-pVQZ basis when compared with the `exact' results, but, in contrast to the behaviour seen for the ionisation energy error, the approximate CBS MCCI value is more likely to have a greater error than the aug-cc-pVQZ results.

\begin{figure}[h!]\centering
\includegraphics[width=.45\textwidth]{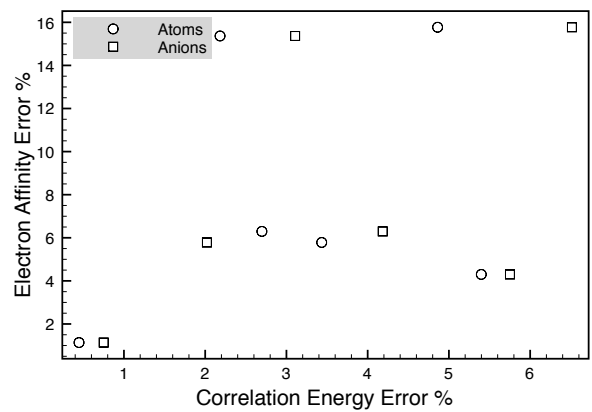}
\caption{Percentage error in the electron affinity plotted against the percentage error in the correlation energy of the atom (circles) or anion (squares). All results are for MCCI at $c_{\text{min}}=5\times10^{-4}$ compared with i-FCIQMC\cite{AlaviFCIQMCAffinity} both using an aug-cc-pVQZ basis. }\label{fig:affinityErrorvEcorr}
\end{figure}

Figure \ref{fig:affinityErrorvEcorr} shows that the percentage errors in the correlation energy are actually fairly similar for the atom and anion with a fixed value of electron affinity error, i.e., within a given system.  It appears that the small differences between the energy of the atom and anion often amplifies the errors for the electron affinity here.  This is in contrast to the ionisation errors where the errors in correlation energies tended to be larger than those of the ionisation energy (Fig.~\ref{fig:affinityErrorvEcorr}).  The large errors in the affinity for boron means that there appears to be less of a linear relationship here compared with the previous results for the ionisation energy: now the statistical correlation for the atom results $0.33$ and the value for the anion results is $0.50$.

\section*{CONCLUSIONS}

In this work we have demonstrated that not only is MCCI useful for energy calculations but that other properties, in the form of multipole moments, may generally be calculated to sufficiently high accuracy with it when compared with FCI results yet using a very small fraction of the FCI space (see table \ref{tbl:multipoles}).  By using an aug-cc-pVDZ basis, resulting in a full configuration space far beyond current FCI, MCCI results could also be seen to generally give a fairly good agreement with experiment.   For the calculations of ground-state multipole moments at equilibrium geometries, methods based on coupled cluster would be expected to be one of the most efficient choices.   However we note that MCCI can perform substantially better when the system moves away from equilibrium and is multireference. In addition the use of different spin states and excited states present no problems, in theory, for MCCI. However we note that the results for excited states appeared to be more sensitive to the removal of states for the dipole of the first excited state of $A_{1}$ symmetry within $C_{2V}$ for carbon monoxide.  Investigations into the use of state-averaging to prevent these oscillations are planned.

We saw that ionisation energies for atoms can be calculated using MCCI with an aug-cc-pVQZ basis to an error of less than ~$1.2\%$ compared with FCIQMC\cite{AlaviFCIQMCIonise} and less than $~2\%$ compared with `exact' non-relativistic results. We note that the largest FCI space was for magnesium with $10^{17}$ SDs and this was not calculated in Ref.~\citenum{AlaviFCIQMCIonise} using FCIQMC due to time constraints, but here only required about $1,000$ CSFs. Similarly to the results of FCIQMC\cite{AlaviFCIQMCIonise,AlaviFCIQMCAffinity} we found that the system rather than just the size of the FCI space was a factor in the cost and accuracy of the a calculation:  oxygen had the largest percentage error compared with FCIQMC here and required $3,541$ CSFs compared with a FCI space of `only' $10^{13}$.  Electron affinity calculations were more challenging for MCCI. Although the absolute errors with i-FCIQMC\cite{AlaviFCIQMCAffinity} were fairly similar to the ionisation energies at less than $10$ milliHartree when using $c_{\text{min}}=5\times10^{-4}$, the percentage error was much higher, partly due to the much smaller energies involved: the largest MCCI error is now around $20\%$ when compared with the exact and $15\%$ when compared with i-FCIQMC.  The highest error with respect to i-FCIQMC was oxygen and this could be reduced to around $8\%$ by lowering $c_{\text{min}}$ to $10^{-4}$ but now $37,225$ CSFs were required for the anion compared with the FCI space of around $10^{13}$.  We note that the percentage error in the MCCI correlation energy at $c_{\text{min}}=5\times10^{-4}$ was fairly similar for a given atom, its cation and its anion.  It was also always lower than $7\%$.  This suggests that MCCI performs similarly for the calculation of ionisation energies and electron affinities but the smaller values of the latter means it has larger percentage errors.   

MCCI appears to possibly be a useful alternative for the calculation of ionisation energies of atoms using a very compact wavefunction, however for electron affinities the larger fraction of the FCI space that appears necessary to be explored for a balanced description of the anion at higher accuracy suggests that other methods may be more appropriate here if consistently small percentage errors are required.  It would appear that for situations where more standard methods have difficulties, such as excited states, then MCCI could be a useful tool for the calculation of properties such as multipoles. The results for multipoles at geometries away from equilibrium were seen to be substantially better at approximating the FCI result when using MCCI than when employing methods based on a single reference suggesting that  MCCI could also be useful for the calculation of multipole surfaces.

\subsection*{ACKNOWLEDGMENTS}
We thank the European Research Council (ERC) for funding under the European Union's Seventh Framework Programme (FP7/2007-2013)/ERC Grant No. 258990.


\providecommand{\noopsort}[1]{}\providecommand{\singleletter}[1]{#1}%
\end{document}